\documentstyle{mn}

\newcounter{parentequation}
\def\beglet{
  \addtocounter{equation}{1}%
  \setcounter{parentequation}{\value{equation}}%
  \setcounter{equation}{0}%
  \def\theequation{\arabic{parentequation}\alph{equation}}%
  \ignorespaces
}
\def\endlet{
  \setcounter{equation}{\value{parentequation}}%
  \def\theequation{\arabic{equation}}%
}

\def\ltsima{$\; \buildrel < \over \sim \;$}
\def\gtsima{$\; \buildrel > \over \sim \;$}
\def\simlt{\lower.5ex\hbox{\ltsima}}
\def\simgt{\lower.5ex\hbox{\gtsima}}

\def\hmpc{h^{-1} {\rm Mpc}}
\def\etal{{\it et al.}\rm}

\begin{document}

\title[Power Spectrum of the APM Survey]
{Maximum Likelihood Estimates of the Two- and Three-Dimensional Power
Spectra of the APM Galaxy Survey}

\author[G. Efstathiou and S.J. Moody]{G. Efstathiou and S.J. Moody \\
Institute of Astronomy, Madingley Road, Cambridge, CB3 OHA.}

\maketitle

\begin{abstract}
We estimate the two- and three-dimensional power spectra, $P_2(K)$ and
$P_3(k)$, of the galaxy distribution by applying a maximum likelihood
estimator to pixel maps of the APM Galaxy Survey. The analysis
provides optimal estimates of the power spectra and of their
covariance matrices if the fluctuations are assumed to be Gaussian.
Our estimates of $P_2(K)$ and $P_3(k)$ are in good agreement with
previous work but we find that the errors at low wavenumbers have been
underestimated in some earlier studies.  If the galaxy power spectrum
is assumed to have the same shape as the mass power spectrum, then the
APM maximum likelihood $P_3(k)$ estimates at $k \le 0.19 h {\rm
Mpc}^{-1}$ constrain the amplitude and shape parameter of a
scale-invariant CDM model to lie within the $2\sigma$ ranges $0.78 \le
(\sigma_8)_g \le 1.18$ and $0.05 \le \Gamma \le 0.38$.  Using the
Galactic extinction estimates of Schlegel, Finkbeiner and Davis, we
show that Galactic obscuration has a negligible effect on galaxy
clustering over most of the area of the APM Galaxy Survey.

\vskip 0.2 truein
\end{abstract}


\section{Introduction}

In this paper we analyse the three-dimensional power spectrum of
galaxy clustering using the APM Galaxy Survey (Maddox \etal~ 1990a, b,
c). The APM Galaxy Survey is a two-dimensional catalogue of galaxies
complete to a magnitude limit of $b_J = 20.5$ and covering an area of
approximately $12$ percent of the sky. The survey has been used to estimate
the angular two-point correlation function  $w(\theta)$ and the angular power
spectrum $P_2(K)$, which are related to their three-dimensional
analogues $\xi(r)$ and $P_3(k)$ via simple integral equations (Limber
1953; Groth and Peebles 1977; Baugh and Efstathiou 1994). Recovering
the three-dimensional power spectrum from angular statistics therefore
requires stable numerical techniques for inverting these integral
equations.

Baugh and Efstathiou (1993) described a technique for recovering the
three-dimensional power spectrum from measurements of the angular
correlation function. The three-dimensional power spectrum was
parameterized by a set of amplitudes $P_3^i$ (or `bandpowers') over
bands of wavenumbers centred at wavenumber $k_i$. The integral
equation relating $w(\theta)$ to $P_3(k)$ was solved using Lucy's
(1974) iterative deconvolution technique. A similar technique was
applied by Baugh and Efstathiou (1994) to recover $P_3(k)$ from
estimates of the two-dimensional power spectrum $P_2(K)$ and by Baugh
(1996) to recover $\xi(r)$ from $w(\theta)$. These investigations show
that  Lucy's algorithm can provide a stable inversion. However, it is
difficult to derive a reliable covariance matrix for the recovered
estimates of $P_3^i$. Baugh and Efstathiou (1993, 1994) derived
estimates of the errors by computing the scatter in the $P_3^i$
derived from four nearly equal areas of the APM survey. However, since
the number of zones is small,  these error estimates are crude and
cannot be used to fit theoretical models with any confidence.

  Recently, Dodelson and Gazta\~naga (2000) have described a method of
inverting $w(\theta)$ to recover $P_3^i$ that employs a Bayesian prior
to contrain the smoothness of the inversion. This method can return a
covariance matrix for $P_3^i$, but requires an estimate of the
covariance matrix of the input estimates of $w(\theta_i)$ and a model
for the Bayesian prior. Eisenstein and Zaldarriaga (1999) present
another inversion technique using singular value decomposition (see
{\it e.g.} Press \etal~ 1992). Their method also recovers the
covariance matrix for $P_3^i$ but requires estimates of $w(\theta)$
and its covariance matrix as inputs.

  The purpose of this paper is two-fold. Firstly, to assess the
effects of Galactic extinction on large scale clustering in the APM
Survey using the extinction model of Schlegel, Finkbeiner and Davis
(1998, hereafter SFD) based on the COBE/DIRBE and IRAS maps.
Secondly, to apply to the APM Survey modern maximum likelihood (ML)
techniques similar to those used to estimate the power spectrum of the
cosmic microwave background (CMB) anisotropies (Bond, Jaffe and Knox
1998; de Bernardis \etal~ 2000; Hannay \etal, 2000). With the increase
in computer power over the ten years since the APM survey was
completed, it is now feasible to perform a direct ML estimate of the
angular power spectrum over wavenumbers extending into the non-linear
regime.  This provides an optimal estimate (under certain assumptions)
of the power spectrum and its covariance matrix in a conceptually
straightfoward way, avoiding the need for estimators of $P_2(K)$ or
$w(\theta)$ that require a model of the true power spectrum. (See {\it
e.g.}  Hamilton, 1997a, b; Tegmark 1997, Kerscher \etal~ 2000, and
references therein for a discussion of estimators of $P_2(K)$ and
$w(\theta)$). An additional advantage of ML methods is that it is as
easy to compute bandpower estimates of the three-dimensional power
spectrum $P_3^i$ (and its covariance matrix) as it is to estimate the
two-dimensional power spectrum. The inversion from two to three
dimensions can therefore be done with the same computer code and
without the need for any assumptions other than that the underlying
fluctuations obey Gaussian statistics.

  The outline of this paper is as follows. Section 2 describes the
method and applies it to Gaussian realizations of the APM Survey. In
Section 3, we use the SFD dust maps to show how the two-dimensional
power spectrum is affected by Galactic extinction. A model for the
mean distribution of galaxies with redshift $dN(z)/dz$ is constructed
using data from the 2dF Galaxy Redshift Survey and this is used to
compute the two- and three-dimensional power spectra by
ML. Constraints on theoretical models are discussed in Section 4 and
our conclusions are summarized in Section 5.

\section{Method}

\subsection{Relations between power spectra and correlation functions}

In this Section we follow the notation of Baugh and Efstathiou (1993,
1994, hereafter refered to as BE93 and BE94).  The angular correlation
function $w(\theta)$ is related to the spatial correlation function 
$\xi(r, t)$ via the relativistic form of Limber's equation
\begin{equation}
w(\theta) = {2 \int_0^\infty \int_0^\infty x^4 F^{-2} a^6 p^2(x) \xi(r, t)
dx du  \over \left [ \int_0^\infty x^2 F^{-1}a^3 p(x) dx \right ]^2},
 \label{M1}
\end{equation}
Peebles (1980, \S50.16). In this equation, $p(x)$ is the selection
function of the survey (the probability that a galaxy at coordinate
distance $x$ is detected in the survey), $a$ is the cosmological scale
factor, and the metric is
\begin{equation}
ds^2 = c^2 dt^2 - a^2 \left[ dx^2/F^2(x) + x^2\;d\theta^2 + x^2 {\rm
 sin}^2\theta\;d\phi^2 \right]. \label{M2}
\end{equation}
Equation (\ref{M1})  assumes that the clustering of galaxies is
independent of luminosity. However, this is quite a weak assumption
for a magnitude limited optical survey since most of the galaxies have
luminosities in a narrow range around the characteristic luminosity
$\cal L^*$ of the Schechter (1976) luminosity function.  The physical
separation between galaxy pairs separated by an angle $\theta$ on the
sky is
\begin{equation}
r = a \left[ u^2/F^2(x) + x^2\theta^2 \right ]^{1/2}, \label{M2a}
\end{equation}
where we have assumed that the angle $\theta$ is small. In the rest of
this paper we adopt a spatially flat cosmological model with matter
density parameter $\Omega_m = 0.3$ and a cosmological constant
contributing $\Omega_\Lambda = 0.7$. 

The spatial correlation function $\xi(r, t)$ is related to the
three-dimensional power spectrum $P_3(k, t)$ by
\begin{equation}
\xi(r, t) = {1 \over 2 \pi^2} \int_0^\infty P_3(k, t) {{\rm sin} (kr/a) \over (kr/a)}
k^2\; dk, \label{M3}
\end{equation}
and following BE93 we will assume that $P_3(k, t)$ is a separable
function of comoving wavenumber $k$ and redshift $z$.
\begin{equation}
 P_3(k, t) = {P_3(k) \over (1 + z)^\alpha}. \label{M4}
\end{equation}

The two-dimensional power spectrum $P_2(K)$ is related to the angular
correlation function by  
\begin{equation}
 P_2(K) = 2 \pi \int_0^\infty w(\theta) J_0(K\theta) \theta d\theta. \label{M5}
\end{equation}
From equations (\ref{M1}), (\ref{M3})--(\ref{M5}), the
two-dimensional power spectrum is related to the three dimensional
power spectrum by  the integral equation
\beglet
\begin{equation}
 K P_2(K) = \int_0^\infty g(K/k) P_3(k)\; dk ,\label{M6}
\end{equation}
where the kernel $g(K/k)$ is 
\begin{equation}
 g(K/k) =  {1 \over {\cal N}^2 \Omega_s^2}
\left [  \left ( {d N \over dz} \right )^2 \left ( {dz \over dx } \right )^2
{F(x) \over (1 + z)^\alpha} \right ]_{x = K/k}. \label{M7}
\end{equation}
\endlet
(see BE94) and  we have written the selection function $p(x)$
in terms of the redshift distribution $dN/dz$ of the sample
\begin{equation}
 p(x)  =  {1 \over {\cal N}^2 \Omega_s^2}
 {F(x) \over x^2 a^2} {dN \over dz} {dz \over dx},  \label{M8}
\end{equation}
where ${\cal N}$ is the mean surface density of galaxies and
$\Omega_s$ is the solid angle of the survey.  If we know the redshift
distribution of a two-dimensional survey, the three-dimensional power
spectrum can be recovered from estimates of the two-dimensional power
spectrum by inverting equation (\ref{M6}) using, for example, Lucy's
(1974) method as described by BE94. However, in the next section we
show that the inversion can be done by using a ML estimator.
The ML method actually solves two problems simultaneously,
solving the inversion probem and providing an optimal estimator
of the power spectra $P_2(K)$ and $P_3(k)$.

\subsection{Maximum likelihood estimator}

Assume that the galaxy catalogue is pixelized into a map 
of $N$ identical pixels with galaxy count $n_i$ in the i'th pixel. 
We define the data vector $\Delta$ as
\begin{equation}
\Delta_i = {n_i - \langle n \rangle
\over \langle n \rangle }, \label{ML1}
\end{equation} 
where $\langle n \rangle$ is the mean galaxy count per pixel.

If we assume that the $\Delta_i$ constitute a Gaussian random field, the
likelihood function is
\beglet
\begin{equation}
 {\cal L}   =  {1 \over (2 \pi)^{N/2} \vert C \vert^{1/2}}
{\rm exp} \left ( - {1 \over 2} \Delta^T C^{-1} \Delta \right),  \label{ML2a}
\end{equation}
where $C$ is the covariance matrix
\begin{equation}
 C_{ij}   =  \left < \Delta_i \Delta_j \right >. \label{ML2b}
\end{equation}
\endlet
From the definition of $\Delta_i$,
\beglet 
\begin{equation}
 C_{ij}   =  {1 \over \langle n \rangle } \delta_{ij} + \bar w (\theta_{ij}),
 \label{ML3a}
\end{equation}
where for square pixels of width $\theta_c$
\begin{eqnarray}
\bar w(\theta_{ij}) & = {1 \over 2 \pi^2} \int_0^\infty \int_0^\pi
P_2(K) K \;{\rm cos}\left (K\theta_{ij} {\rm cos} \phi \right ) \qquad  \nonumber \\
& \times W_c^2(K\theta_c, \phi) d\phi dK,  \label{ML3b}
\end{eqnarray}
and
\begin{equation}
W_c(K\theta_c, \phi) = 
 {\rm sinc}\left ( {K \theta_c \over 2} {\rm cos} \phi \right )
\;{\rm sinc} \left ( {K \theta_c \over 2} {\rm sin} \phi \right ).
\end{equation}
\endlet
For angular separations much greater than the pixel size, equation (\ref{ML3b})
simplifies to.
\begin{equation}
\bar w(\theta_{ij}) \approx {1 \over 2 \pi} \int_0^\infty P_2(K) K J_0(K \theta_{ij}) dK ,
\qquad \theta \gg \theta_c. \label{ML4}
\end{equation}
Equations (\ref{ML3b})
and (\ref{ML4}) have been derived in the small angle limit $\theta_{j} \ll 1$,
which is a good approximation for the APM Galaxy Survey. This assumption
is easily dropped, however, in which case equation (\ref{ML4}) reads
\begin{eqnarray}
\bar w(\theta_{ij}) &= {1 \over 4 \pi} \sum_\ell (2 \ell + 1) P_2(\ell) P_\ell
({\rm cos} \theta_{ij}) . \label{ML5}
\end{eqnarray}
In analogy with analyses of cosmic microwave background
 anisotropies, the angular wavenumber $K$
is equivalent to the multipole moment $\ell$ and the angular 
power spectrum $P_2(K)$ is equivalent to $C_\ell$ (see {\it e.g.} Bond
and Efstathiou 1987).

\begin{figure*}
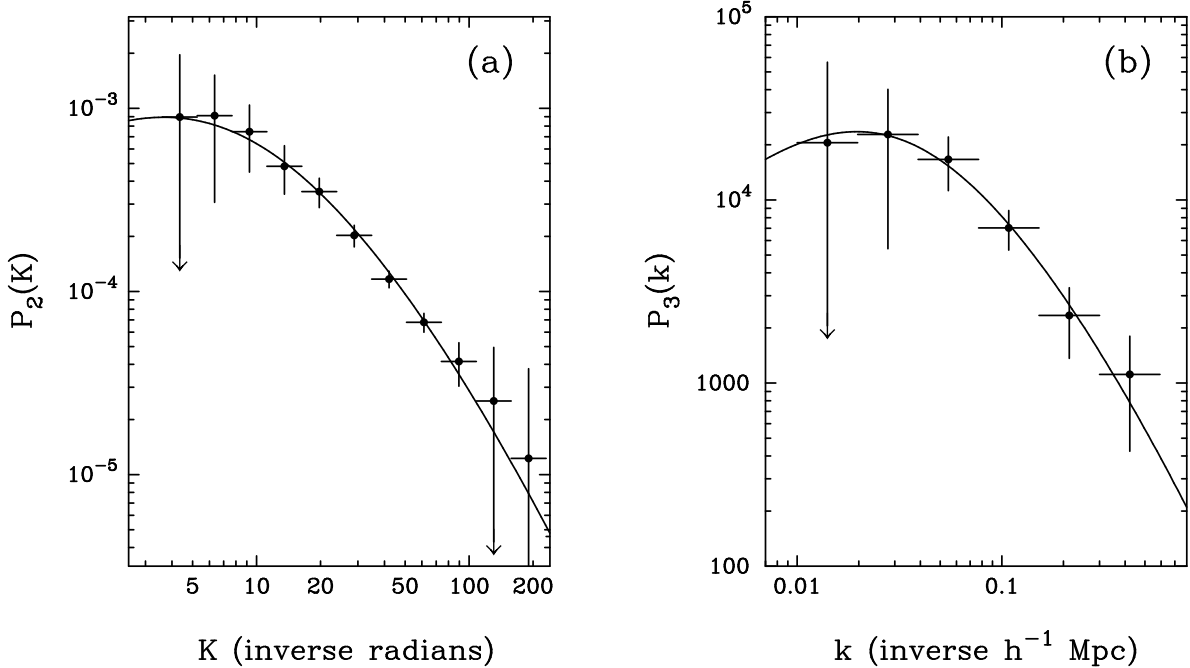


\vskip 3.8 truein

\includegraphics{pgfig1a.ps}
\includegraphics{pgfig1b.ps}

\caption
{ML bandpower estimates of the power spectra
determined from simulated Gaussian random fields generated with a CDM
power spectrum with shape parameter $\Gamma = 0.2$. The simulations
mimic a square patch of the sky of side $100^\circ$ with the selection
function of the APM Galaxy Survey limited to $b_J = 19.5$. The solid
lines show the input two- and three-dimensional power spectra. The
points show the mean bandpower estimates computed from 40 maps each
containing $1024$ pixels. The vertical error bars show $1\sigma$ error
estimates for a single simulation derived from the Fisher matrix. The
horizontal error bars show the widths of each band.}
\label{figure1}
\end{figure*}

Following Bond, Jaffe and Knox (1998), the likelihood function (\ref{ML2a})
can be maximized iteratively with respect to a set of parameters $a_p$.
Starting from an initial guess for the $a_p$, the changes in the parameters
$\delta a_p$ at each iteration are calculated from
\begin{equation}
 \delta a_p = {1 \over 2} \sum_{p^\prime}  F_{p p^\prime}^{-1}
{\rm Tr} \left [ ( \Delta \Delta^T - C) \left
( C^{-1} {\partial C \over \partial
a_{p^{\prime}}} C^{-1} \right ) \right ],
 \label{ML6}
\end{equation}
where $F_{p p^\prime}$ is the Fisher matrix
\begin{equation}
 F_{p p^\prime} =  - \left < {\partial^2 {\rm ln} {\cal L}
\over \partial a_p \partial a_{p^\prime} } \right >   
   = {1 \over 2} {\rm Tr} \left [ C^{-1} {\partial C
\over \partial a_p} C^{-1} { \partial C \over \partial a_{p^\prime}} 
\right ].
 \label{ML7}
\end{equation}

The parameters $a_p$ can be chosen to be bandpower estimates of
the two-dimensional power spectrum $P_2(K)$ or of the three-dimensional
power-spectrum $P_3(k)$. For these cases, the angular correlation function
in equation (\ref{ML3a}) is computed from the sum
\begin{equation}
\bar w(\theta_{ij})  =  \sum_p a_p \beta_p(\theta_{ij}),  \label{ML8a}
\end{equation}
where
$$
 \beta_p = \left\{   \begin{array}{ll} 
{1 \over 2 \pi^2}\int_{K_p}^{K_{p+1}}\int_0^\pi
K \;{\rm cos}\left (K\theta_{ij} {\rm cos} \phi \right ) \\
\qquad \times  W_c^2(K\theta_c, \phi) \;d\phi dK, 
 & \mbox{for $P_2(K)$,}   \\ 
{1 \over 2 \pi^2} \int_0^\infty\int_{k_p}^{k_{p+1}}\int_0^\pi
g(K/k)\;{\rm cos}\left (K\theta_{ij} {\rm cos} \phi \right ) \\
\qquad \times  W_c^2(K\theta_c, \phi) \;d\phi dk dK, & \mbox{for $P_3(k)$.} 
\end{array} \right.   
$$
These integrals depend only on the binning of the parameters and on
the pixel scale,  so they can be computed once and stored. The
computing time required to find the ML is dominated by
the computation of the inverse matrix $C^{-1}$ and the multiplication
of $N \times N$ matrices (both of which scale as $N^3$). Our
implementation on an SGI Origin 200 workstation takes a few hours to
converge to a solution for $N\approx 4000$.

\subsection{Tests of the Method}

We have tested the algorithm on simulated two-dimensional Gaussian
random fields. We assume that the three-dimensional power spectrum
is that of a linear adiabatic scale-invariant CDM model with a shape
parameter of $\Gamma = 0.2$ in the parameterization of Efstathiou,
Bond and White (1992).  The two-dimensional power spectrum was
computed from equation (\ref{M6}) using a model for the redshift
distribution of the APM Survey limited at $b_J=19.5$ (see Section
3.2 below). We adopt an evolution parameter of $\alpha=0$ and
normalize the spectra so that the rms fluctuation amplitude of the
galaxy distribution averaged in spheres of radius $8 \hmpc$ spheres,
$(\sigma_8)_g$, is unity. We used an $1024^2$ FFT to generate a
periodic Gaussian density field from the two-dimensional power
spectrum in a $400^\circ \times 400^\circ$ square from which we
selected a $100^\circ \times 100^\circ$ patch regridded into $32\times
32$ pixels for input into the ML code. The pixel size of the input
catalogues is therefore $3.12^\circ$,  but they include small
scale power because they were generated on a grid of
much  higher resolution.
 
The ML reconstructions averaged over $40$ simulations are shown in
Figure 1. Convergence to the ML solution for both the two- and
three-dimensional power spectra is usually achieved within 5--10
iterations. The error bars shown on the points are computed from the
inverse of the Fisher matrix, $\sigma_i^2 = F_{ii}^{-1}$, and are in
excellent agreement with the scatter between simulations.

There are a few subtle points about the analysis worth some discussion:

\smallskip

\noindent
[1] The sums over the bandpower parameters $a_p$ in equation
(\ref{ML8a}) are performed over a finite range of wavenumber $K_{min}
< K < K_{max}$ (or $k_{min} < k < k_{max}$, depending on whether we
are estimating the two- or three-dimensional power spectra).  Ignoring
power from wavenumbers outside these ranges leads to small biases in
the ML solutions.  In the examples shown in Figure 1, we have
explicitly included integrals over the power spectra at $K < K_{min}$
and $K > K_{max}$ assuming the input target power spectrum which is,
of course, known. This removes any biases at large and small
wavenumbers as shown in Figure 1. In application to real data, the
power spectrum is unknown. In this case, one can simply increase the
number of parameters extending the range of $K_{min}$ and $K_{max}$
and marginalize over a small number (one should suffice)
of parameters at either end of
the wavenumber range. The remaining parameters will then be free of
any biases.

\smallskip

\noindent
[2] The pixel scale of the maps used to generate Figure 1 corresponds
to a wavenumber $K = 2 \pi/\theta_c \approx 125$. Nevertheless, by
correctly including the window function of the pixels in the integral
of equation (\ref{ML3b}), the power spectrum can be recovered free of
bias on sub-pixel scales, but obviously the errors become large as the
estimates are extended below the pixel scale. In our application to
the APM Survey, the limit on the pixel size is set by size of the data
vector that can be analyzed in a reasonable amount of computer time.
We find that it is possible to analyze maps with pixels of size
$\theta_c = 0.89^\circ$ ($N \approx 4000$ pixels) easily using
workstations.  It would be possible to increase the number of pixels
by using supercomputers and by using Monte-Carlo methods as described
by Oh, Spergel and Hinshaw (1999). However, in the ML analysis it is
assumed that the underlying density fluctuations are Gaussian, whereas
the galaxy distribution is observed to be strongly non-Gaussian on
small scales where the distribution is also non-linear. At magnitude
limits of $b_J \approx 19.5-20.0 $, the angular scales of significant
non-Gaussianity and non-linearity in the APM survey are at $K \simgt
200$. The ML estimator is therefore not guaranteed to be optimal or
even unbiased at wavenumbers higher than $K \simgt 200$. This differs
from the case of applying ML to the CMB anisotropies, where the
assumption of Gaussian fluctuations is physically reasonable for
primary anisotropies on all angular scales.

\smallskip

\noindent
[3] The numerical inversion of an integral equation such as (\ref{M6})
is unstable; the inverted $P_3(k)$ can show wild fluctuations as the
number of bandpowers is increased (see {\it e.g.} BE93, BE94; Dodelson
and Gazta\~naga 2000). The ML method described here imposes no
constraints on the bandpower estimates $a_p$ and so there is no
guarantee that the recovered power spectra will be smooth. As the
number of bandpowers is increased, the ML solutions (particularly for
$P_3(k)$) will begin to show oscillations. However, if we fit a
theoretical model characterized by a few parameters to the data {\it
using the full covariance matrix of the estimates} (see Section 4),
then the best fitting parameters will be insensitive to the number of
bandpowers and to oscillations in $P_3(k)$.

\smallskip

\noindent
[4] In analyzing the simulations, the mean galaxy count per pixel was
estimated from each map by computing
\begin{equation}
 \langle n \rangle = {1 \over N} \sum_i n_i. \label{ML9}
\end{equation}
This is not strictly correct, since $\langle n \rangle$ is the mean
galaxy count averaged over an ensemble of catalogues not the mean
pixel count of a single map. This can introduce a bias that is related to
the `integral constraint' bias in estimates of $w(\theta)$ ({\it e.g.}
Groth and Peebles 1977) and the power spectrum ({\it e.g.} Tadros and
Efstathiou 1996).  More correctly, the mean galaxy count should be
treated as a parameter in the likelihood analysis. Maximising the
likelihood (\ref{ML2a}) with respect to $\langle n \rangle$ gives
\begin{equation}
 \langle n \rangle = {\sum_{ij} n_i n_j C_{ij}^{-1}\over \sum_{ij} n_i
C_{ij}^{-1}}, \label{ML10}
\end{equation}
and so depends on the ML solution for the power spectrum.
In practice the APM Galaxy Survey covers a large enough area that any
bias introduced in using equation (\ref{ML9}) is negligible.

\begin{figure*}

\vskip 7.5 truein
\includegraphics{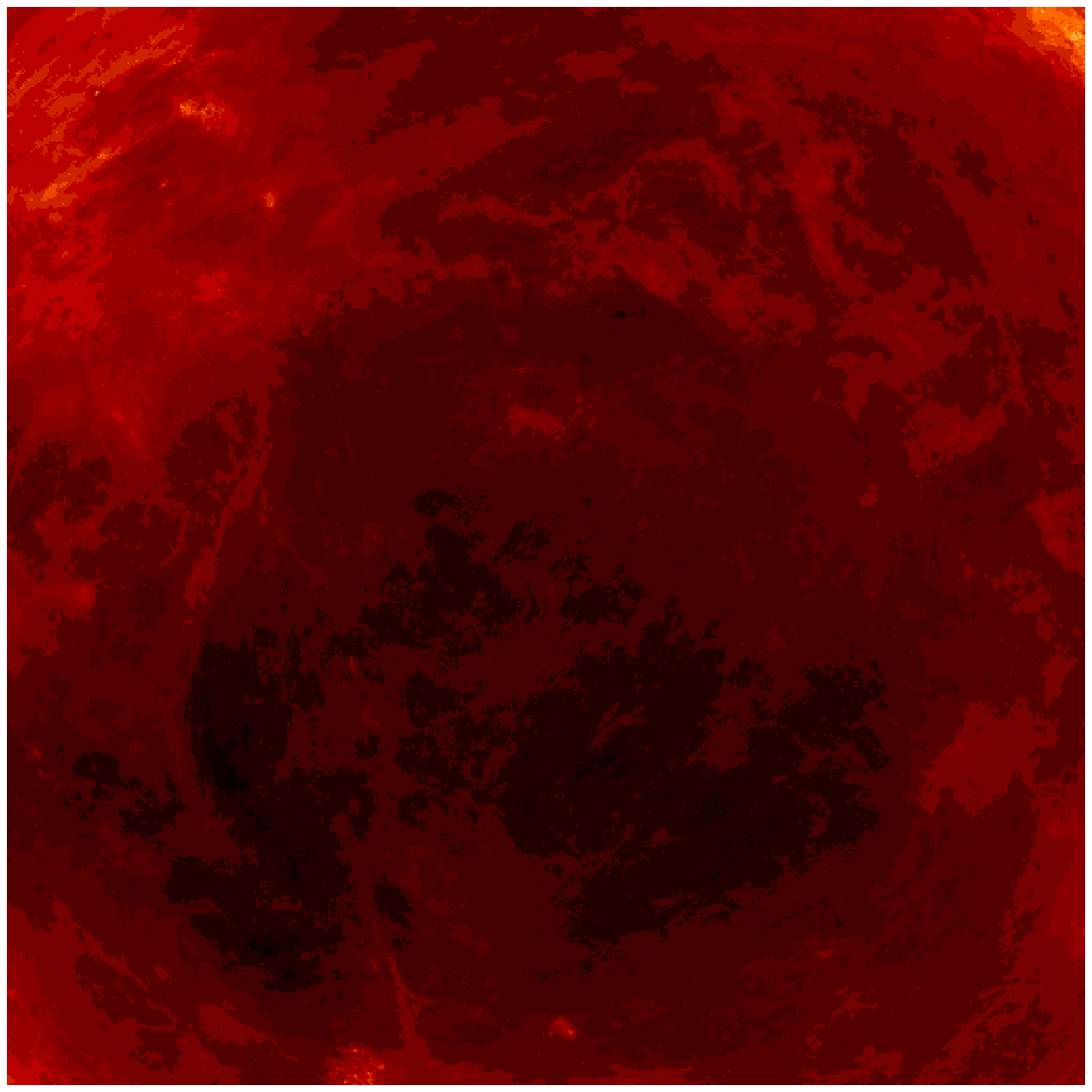}
\includegraphics{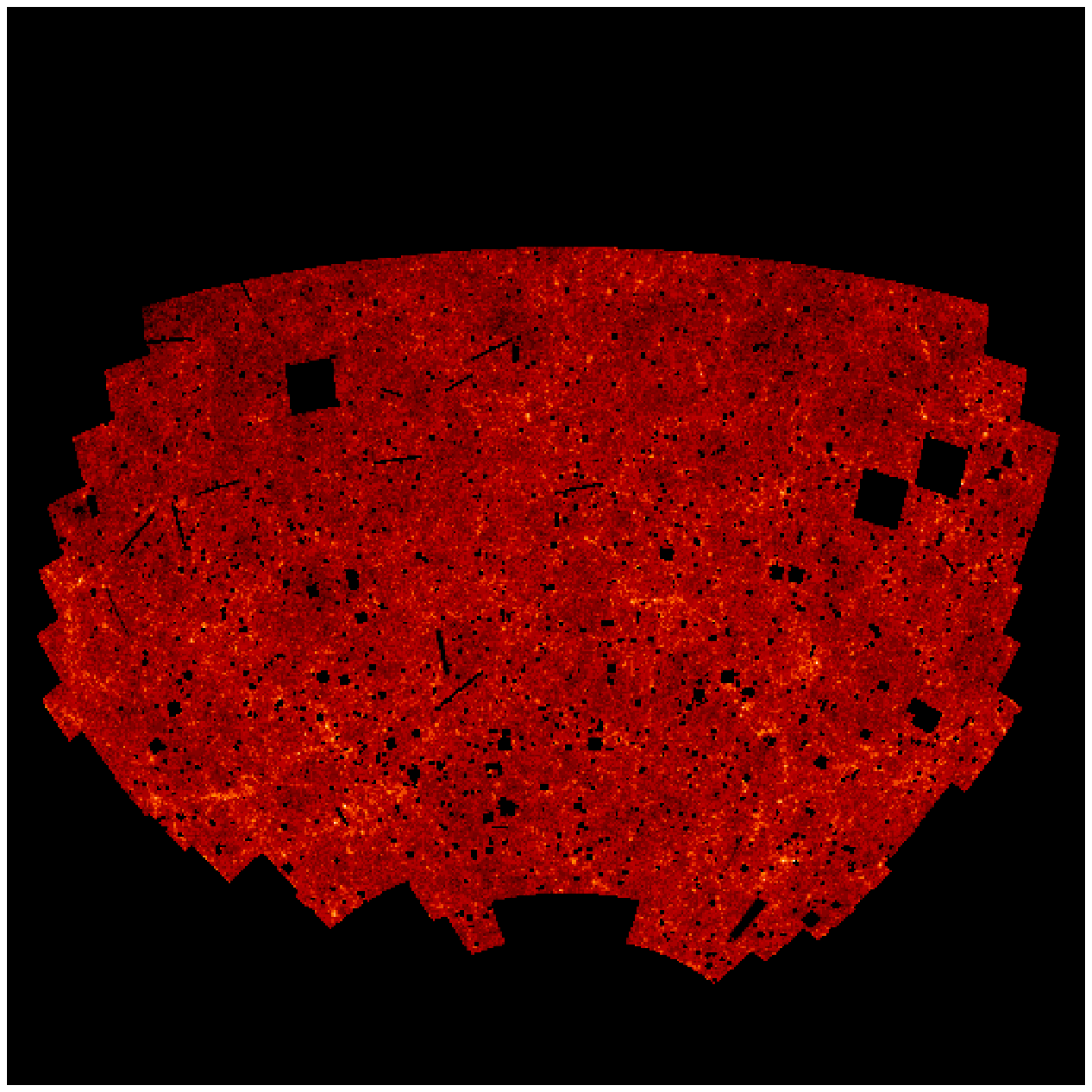}
\includegraphics{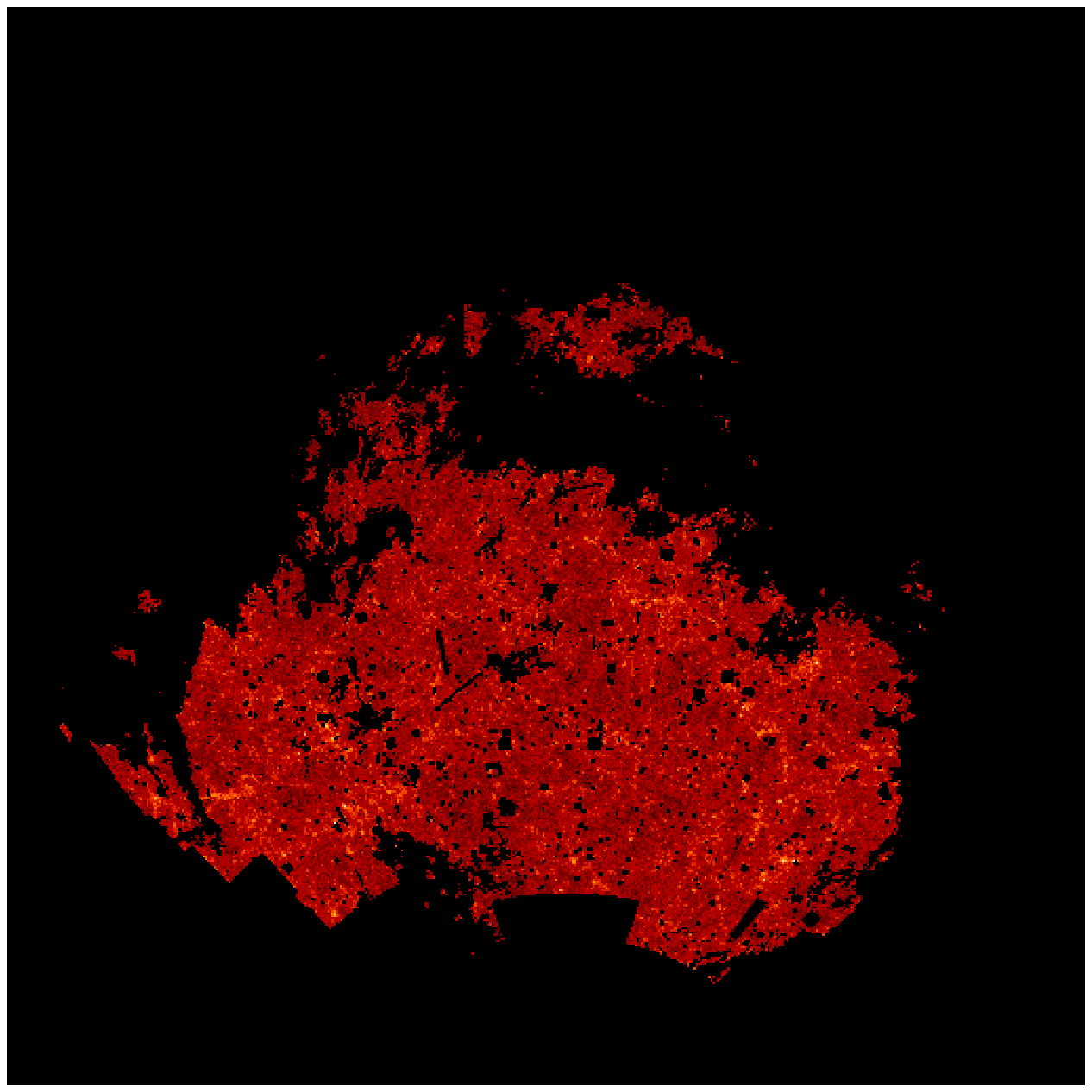}
\includegraphics{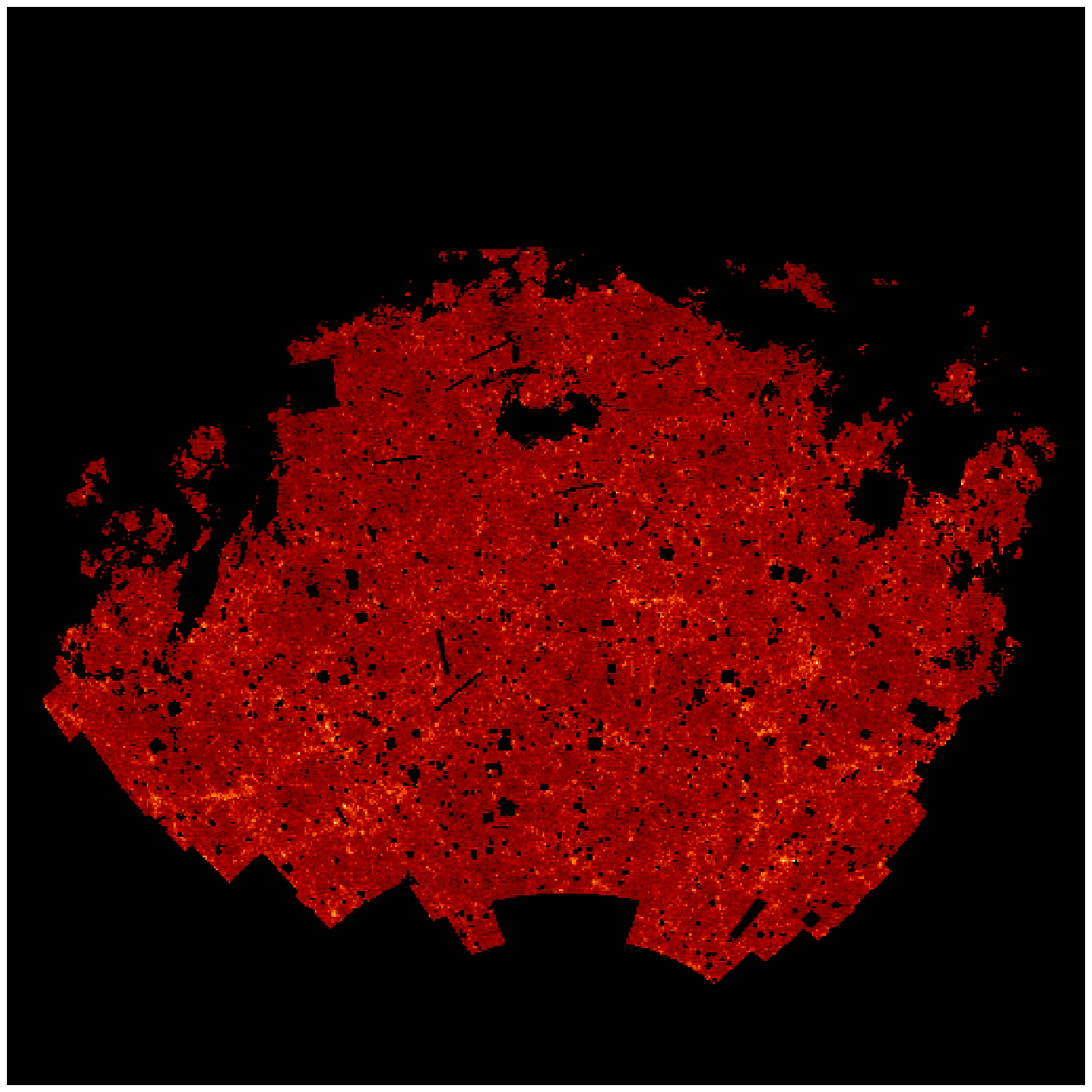}

\caption
{Equal area projections of Galactic extinction and various galaxy
 maps.  Clockwise from the top left: Extinction map in the southern
 Galactic cap computed from SFD; the extended APM Galaxy Survey
 limited at $b_J=20$; the APM Survey in the region with dust
 extinction of less than $0.2$ magnitudes; the APM Survey in the
 region with dust extinction of less than $0.1$ magnitudes.}
\label{figure2}
\end{figure*}

\section{Application to the APM Galaxy Survey}

In Section 3.1, we discuss the effects of Galactic extinction in the
APM Galaxy Survey using the SFD dust map.  This allows us to delineate
an area of the APM Survey in which extinction has a negligible effect
on the power spectrum. In Section 3.2, we use results from a small
subset of the 2dF Galaxy Redshift Survey (see {\it e.g.} Colless,
1999) to derive a model for the redshift distribution of the APM
Survey, improving on the model used by BE93, BE94.  Results from the
ML method are presented in Section 3.3.

\subsection{Input APM Galaxy Catalogue}

The APM Galaxy Survey is described in detail in a series of papers by
Maddox \etal~ (1990a, b, c; 1996). The first version of the catalogue
was based on $185$ UKSTU\footnote{United Kingdom Schmidt Telescope
Unit.}  plates with centres $\delta < -20^\circ$ at high Galactic
latitude in the southern Galactic cap. The survey has since been
extended to include the equatorial region between $-17.5^\circ <
\delta < 2.5^\circ$ and also to include equatorial regions in the
northern hemisphere. Only the southern catalogue, as plotted in Figure
2, is used in this paper. Detailed analyses of the plate matching
algorithm, plate matching errors, completeness, star-galaxy separation
and other possible sources of systematic errors are presented by
Maddox \etal~(1990 b,c; 1996). The survey is largely complete to
$b_J=20.5$, though there are detectable systematic errors (of low
amplitude) in the faint magnitude slice $20< b_J < 20.5$.

\begin{figure*}
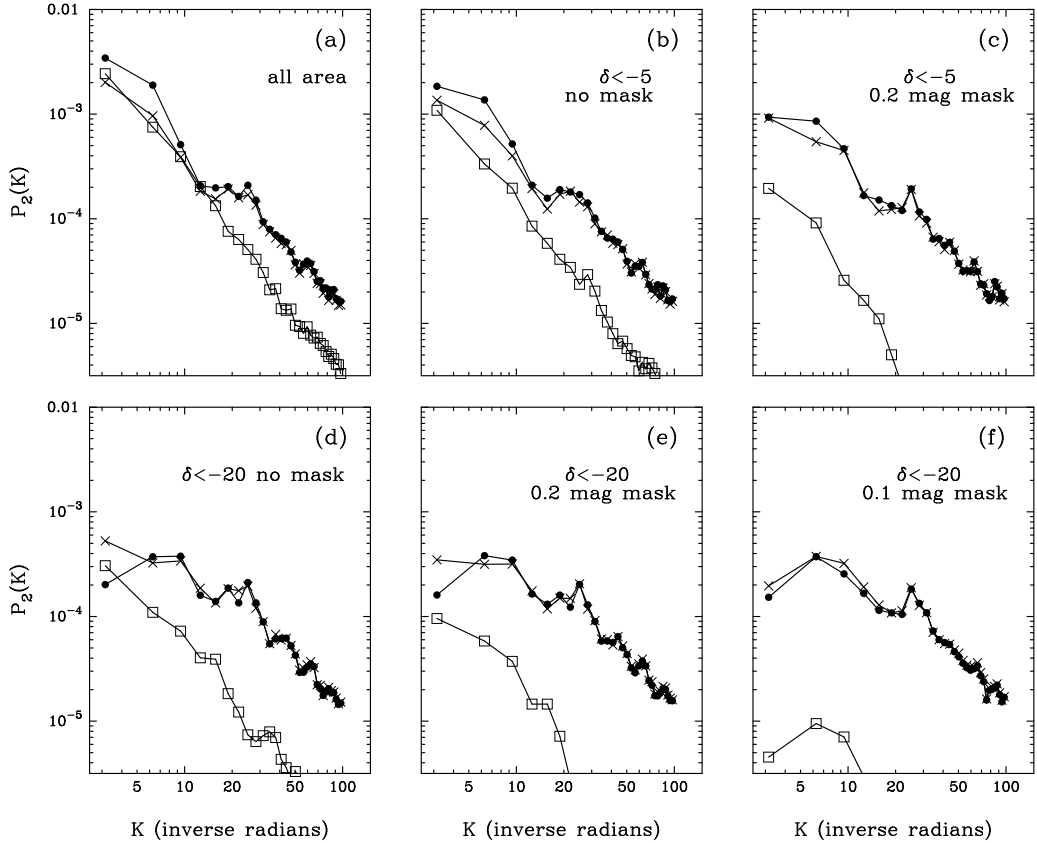


\vskip 4.7 truein
\includegraphics{pgdusta.ps}
\includegraphics{pgdustb.ps}
\includegraphics{pgdustc.ps}

\includegraphics{pgdustd.ps}
\includegraphics{pgduste.ps}
\includegraphics{pgdustf.ps}

\caption
{FFT power spectra for various masked cuts of the APM survey and of
the SFD extinction map . The filled circles show power spectra for the APM
Survey limited at $b_J=20.0$. The crosses show power spectra of APM
galaxies within the same area of sky but at an extinction corrected
magnitude limit of $b^{corr}_J = 20$. The open squares show the power
spectra of the SFD extinction maps over the same area (converted into
a galaxy surface density modulation by equation \ref{A1}). The Figures
are for the following areas: (a) entire APM survey area as plotted in
Figure 2; (b) APM area with $\delta < -5^\circ$; (c) APM area with
$\delta < -5^\circ$ and extinction of $\le 0.2$ mag.; (d) APM area
with $\delta < -20^\circ$, (e) APM area with $\delta < -20^\circ$ and
extinction of $\le 0.2$ mag.; (f) APM area with $\delta < -20^\circ$
and extinction of $\le 0.1$ mag. }
\label{figure3}
\end{figure*}

  SFD have used the COBE/DIRBE and destriped IRAS maps to derive a
map of the dust column density and hence of Galactic extinction. The
Johnson $B$ and $V$ passbands are related to the APM $b_J$ passband by
\begin{eqnarray*}
b_J &\approx B - 0.28(B-V),
\end{eqnarray*}
(see Maddox \etal~ 1990c). The SFD maps of E(B-V)  can therefore be
converted into extinction in the $b_J$ passband by multiplying by a
factor of $4.035$. The extinction computed from the SFD maps in the
region of the southern Galactic pole (SGP) is plotted in Figure 2.
The two plots in the lower panels of Figure 2 show regions of the APM
Survey in which the extinction computed from the SFD is less than
$0.2$ and $0.1$ magnitudes. Evidently, Galactic extinction is
relatively uniform and less than $0.2$ magnitudes over most of the
area of the APM survey at $\delta < -20^\circ$. Regions of extinction
higher than $0.2$ magnitude are confined mainly to the corners at the
top right and left of the APM map.

 Figure 3 shows dust and galaxy power spectra for various 
subsets of the APM area. The power spectra in these figures were
computed from an equal area projection, as in Figure 2, pixelized
into $64 \times 64$ square pixels and applying an FFT to compute
$P_2(K)$ using the estimator of equation (\ref{A5}) below. (These FFT
estimates are  not optimal, but can be computed very
quickly. A comparison of the FFT and ML estimators
is presented in Section 3.3.) In each panel of Figure 3
we show power spectra for the APM Survey galaxies within the specified
area limited at $b_J=20$ (filled circles). The crosses show power spectra
for galaxies within the same region of sky, but with an extinction corrected
magnitude, 
\begin{eqnarray*}
b_J^{corr} = b_J - 4.035E(B-V), 
\end{eqnarray*}
limited to  $b_J^{corr} \le 20$. (Hence the maps are regenerated by applying an
extinction correction to each galaxy). 
The open squares show the power spectrum of the
SFD extinction map, which we have converted into modulations in the 
galaxy surface density in each pixel using
\begin{equation}
n^{ext}_i = \langle n \rangle 10^{-0.45(4.035E(B-V)_i)}. \label{A1}
\end{equation}
Equation  (\ref{A1}) uses an approximate slope for the APM number counts
(see Maddox \etal, 1990d). (Note that the mean extinction
$E(B-V)_i$ is computed by averaging the values over a regular
grid of $16 \times 16$ values within each pixel, to reduce the effects
of small scale variations in the extinction).

 Figure 3 illustrates clearly the effects of galactic
extinction. Figure 3(a) shows that if we use the full APM survey area,
Galactic extinction dominates the power in the APM Survey at
wavenumbers $K \simlt 10$.  Correcting the APM magnitudes for Galactic
extinction results in a small reduction of the power at wavenumbers $K
\simlt 10$, but does not depress the power to the levels seen in
Figures 3(c) - 3(f) for the extinction masked APM maps. There are a
number of possible reasons for this. The conversion from $E(B-V)$ to
extinction in the $b_J$ passband may be wrong. We have tested for this
by correlating the galaxy counts in the pixelized maps with
$E(B-V)$. This is plotted in Figure 4. However, as noted by SFD, at
the limiting magnitude of the APM Survey, the fluctuations in the
number counts caused by galaxy clustering introduce a large
dispersion, so it is difficult to disentangle the effects of Galactic
extinction from galaxy clustering. The general trend of the counts is
consistent with an extinction correction of $4.035E(B-V)$ in the $b_J$
passband, but the correction is not well constrained at high
extinctions.  There may be other sources of gradients in the APM
counts that are uncorrelated or anti-correlated with Galactic
extinction, and so are not removed by the extinction correction. One
effect, noted by Maddox \etal~ (1996) is contamination by stars
(mainly star-galaxy mergers misclassified as galaxies) at low Galactic
latitudes. This effect increases the counts in regions of high
extinction, partially counteracting the effects of obscuration.

\begin{figure}

\vskip 2.8 truein

\includegraphics{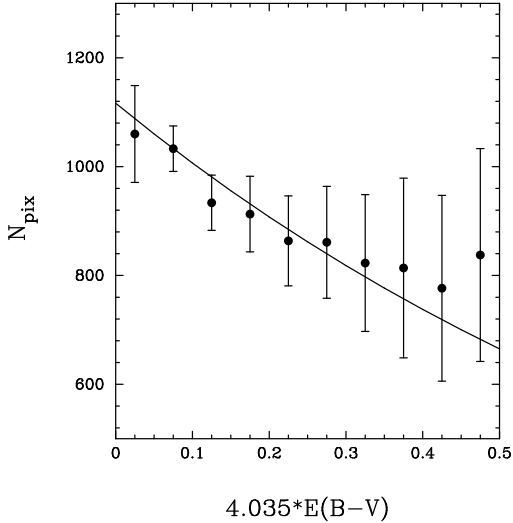}

\caption
{Mean galaxy counts in $1.79^\circ \times 1.79^\circ$ pixels plotted
as a function of $E(B-V)$ computed from the SFD maps. The error bars
show the dispersion of the counts.  The solid line shows
equation \ref{A1}.}
\label{figure4}
\end{figure}

In any case, simply eliminating the most highly obscured parts of the APM
area has a dramatic effect on the power spectrum. As Figures 2 and 3
show, most of the high extinction regions are at $\delta > -20^\circ$
(which is why the original APM Survey area of Maddox \etal~ (1990a-c)
was limited at this declination limit). Galactic extinction within
this area has a negligible effect on the power spectrum except
possible at wavenumber $K \simlt 3$. If all pixels with $\delta >
-20^\circ$ and extinctions of $> 0.2$ magnitudes are removed (Fig 3e),
then the power spectrum of the extinction map is negligible at all
wavenumbers. The power spectrum of the extinction map is reduced still
further (by about an order of magnitude) by removing pixels with an
extinction of $>0.1$ magnitudes (Fig 3f), whereas the power spectrum
of the galaxy distribution hardly changes from that shown in Figures
3d-3f. This is powerful evidence that the power spectrum of the
galaxy distribution in the region $\delta < -20^\circ$ is unaffected
by Galactic extinction. In the rest of this paper, we will analyse
the $\delta < -20^\circ$ map with a $0.2$ magnitude extinction mask
applied as in Figure 3e.

\begin{figure}

\vskip 1.8 truein

\includegraphics{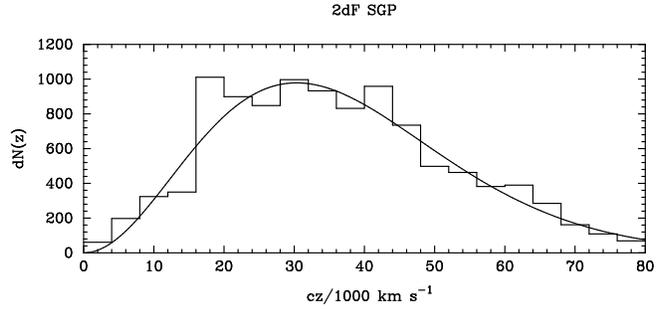}

\caption
{The redshift distribution of galaxies in a high completeness subset
of the 2dF Galaxy Redshift Survey. This sample is
limited at an extinction corrected magnitude limit of $(b_J)_{\rm corr} = 
19.45$. The solid line shows the fit to the dN(z) distribution given
by equation (20).}
\label{figure5}
\end{figure}

\subsection{Redshift distribution}

At the time that the BE93 and BE94 papers were written, few redshifts
had been measured for faint galaxies in the APM Survey. These authors
used the Stromlo/APM redshift survey at bright magnitudes $b_J<17$
(Loveday \etal, 1992) and the small, but deep, pencil beam surveys of
Broadhurst, Ellis and Shanks (1988) and Colless \etal~ (1990, 1993) at
$b_J > 20$ to derive an interpolation formula for the redshift
distribution between these magnitude limits. Here we have used a
subset of $11120$ galaxies in high completeness ($> 0.85$) regions in
the SGP area measured as part of the 2dF Galaxy Redshift Survey
(2dFGRS, see Colless 1999; Folkes \etal~ 1999). The 2dFGRS uses the
APM Galaxy Survey as the source photometric catalogue and has an
extinction corrected (based on the SFD extinction maps) magnitude
limit of $(b_J)_{\rm corr} = 19.45$.

\begin{figure*}
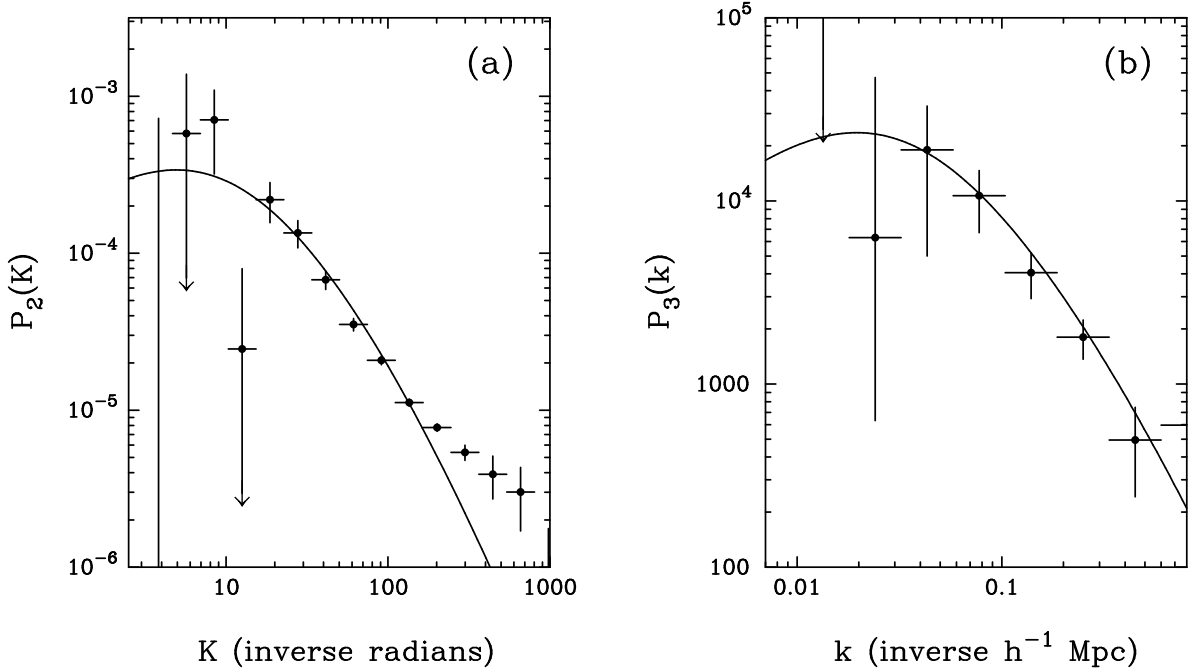


\vskip 3.8 truein

\includegraphics{pgp2k.ps}
\includegraphics{pgp3k.ps}

\caption
{ML bandpower estimates of the two- and three-dimensional power
spectrum of the APM survey limited at $b_J=20.0$. A declination limit
of $\delta = -20^\circ$ and $0.2$ magnitude extinction mask was
imposed.  The error bars show $1\sigma$ error estimates derived from
the Fisher matrix. The solid line shows a scale invariant linear
$\Gamma=0.2$ CDM model normalized so that $(\sigma_8)_g=1.0$.}
\label{figure6}
\end{figure*}

The redshift distribution for this sample is plotted in Figure 5.
We have fitted the redshift distribution by least squares to a form similar
to that used by BE93
\beglet
\begin{equation}
 {dN(z) \over dz} dz =  Az^2 {\rm exp}\left \{ - \left ( {z \over z_c}
\right)^\beta \right \} dz. \label{A2a}
\end{equation}
The best fitting parameters are
\begin{equation}
 z_c = 0.086, \qquad \beta = 1.55, \qquad (b_J = 19.5), \label{A2b}
\end{equation}
where we have used the mean extinction correction of $0.05$ magnitudes
for the 2dFGRS galaxies to convert to uncorrected $b_J$ magnitudes.
The fit of equation (20) is shown by the solid line in Figure 5. The parameters
are  quite close to those used by BE93 for $b_J=19.5$. To extrapolate to
fainter and brighter magnitudes we adopt equation (\ref{A2a}) with $
\beta = 1.55$ and adjust the parameter $z_c$ so that the median
redshift, $z_m$, varies with magnitude limit according to 
\endlet
\begin{equation}
 z_m = 1.36z_c = 0.018(b_J - 17)^{1.5} + 0.046. \label{A3}
\end{equation}
This formula provides an excellent fit to the median redshifts of
published redshift surveys in the magnitude range $17 \le b_J \le 21$
and to the median redshift predicted from fitting the luminosity
function of the 2dFGRS survey. We will use equations
(\ref{A2a}) and (\ref{A3}) to evaluate the kernel $g(K/k)$
of equation ({\ref{M7}).

\subsection{Maximum likelihood power spectra of the APM Survey}

In this Section we show results for the maximim likelihood power
spectra for the APM survey limited at $b_J=20.0$ with a
declination limit of $\delta = -20^\circ$ and a $0.2$ magnitude
extinction mask applied. The input maps covering the area shown in
Figure 2 were generated with $128 \times 128$ pixels of which $4142$
are `active' ({\it i.e.}  correspond to unmasked regions of the
map). In computing the kernel $g(K/k)$ we have assumed that the
evolution parameter $\alpha = 0$.  The parameter $\alpha$ is not known
{\it a priori} and so uncertainties in its value will translate into a
small residual uncertainty in the amplitude of the recovered
three-dimensional power spectrum $P(k)$, though not in its shape (see
BE93; Scranton and Dodelson 2000). We can view the ML
solution with $\alpha=0$ as recovering the three-dimensional power
spectrum at (approximately) the median redshift of the survey, which
for $b_J=20$ is $z_m \approx 0.14$ (equation \ref{A3}).

The results for the two- and three-dimensional power spectra are shown
in Figure \ref{figure6}, together with one $\sigma$ error bars
computed from the Fisher matrices. In Figure 7, we compare these
results with those for the APM Survey limited at $b_J=19.5$ and with
the same sky mask.  The 2d power spectrum for the $b_J=19.5$ map has a
slightly higher amplitude than the $b_J=20$ estimates and is displaced
slightly to higher wavenumbers. This is expected from the scaling
properties of the 2d power spectrum with limiting magnitude (see
BE94). The 3d power spectra for the $b_J=19.5$ and $b_J=20$ maps are
plotted in Figure (7b) and are consistent with each other.

\begin{figure*}
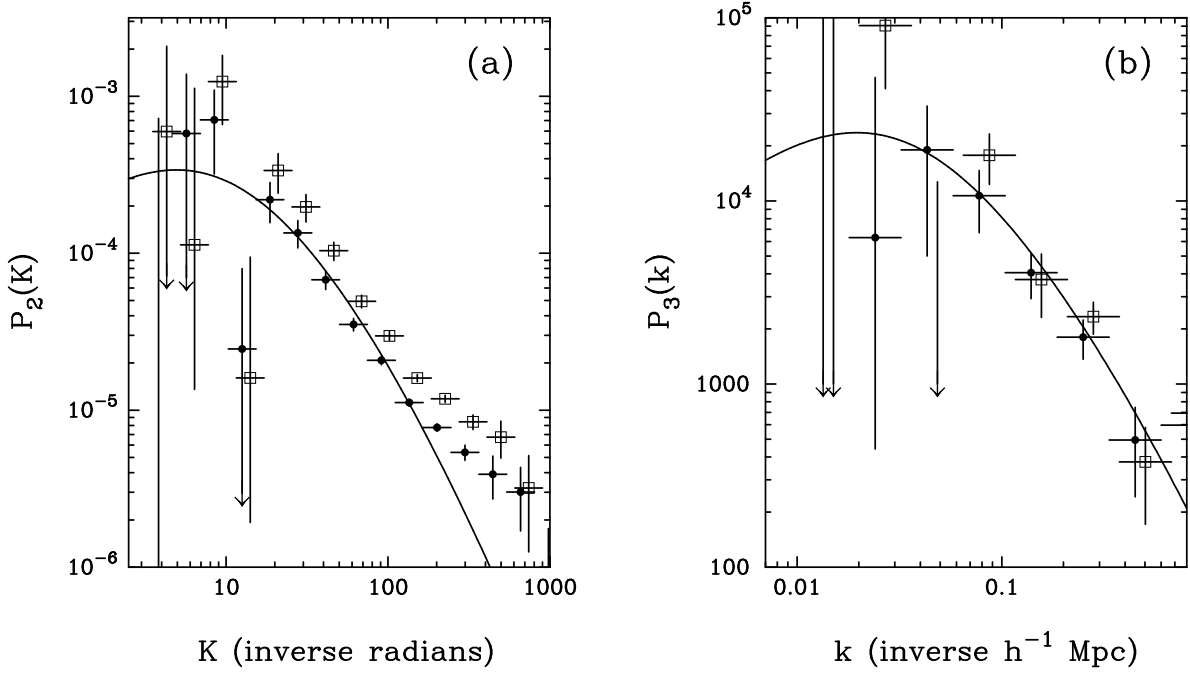


\vskip 3.8 truein

\includegraphics{pgp2k_19.5.ps}
\includegraphics{pgp3k_19.5.ps}

\caption
{ML bandpower estimates of the two- and three-dimensional power
spectrum of the APM Survey. The filled circles show results for the
APM Survey limited at $b_J=20$ as plotted in Figure 6. The open
squares show results for a magnitude limit of $b_J=19.5$.  For
clarity, the $b_J=19.5$ points have been displaced to the right by
$0.05$ in the log. The lines in these figures show a linear
$\Gamma=0.2$ CDM model as in Figure 6.}
\label{figure7}
\end{figure*}

\begin{figure*}
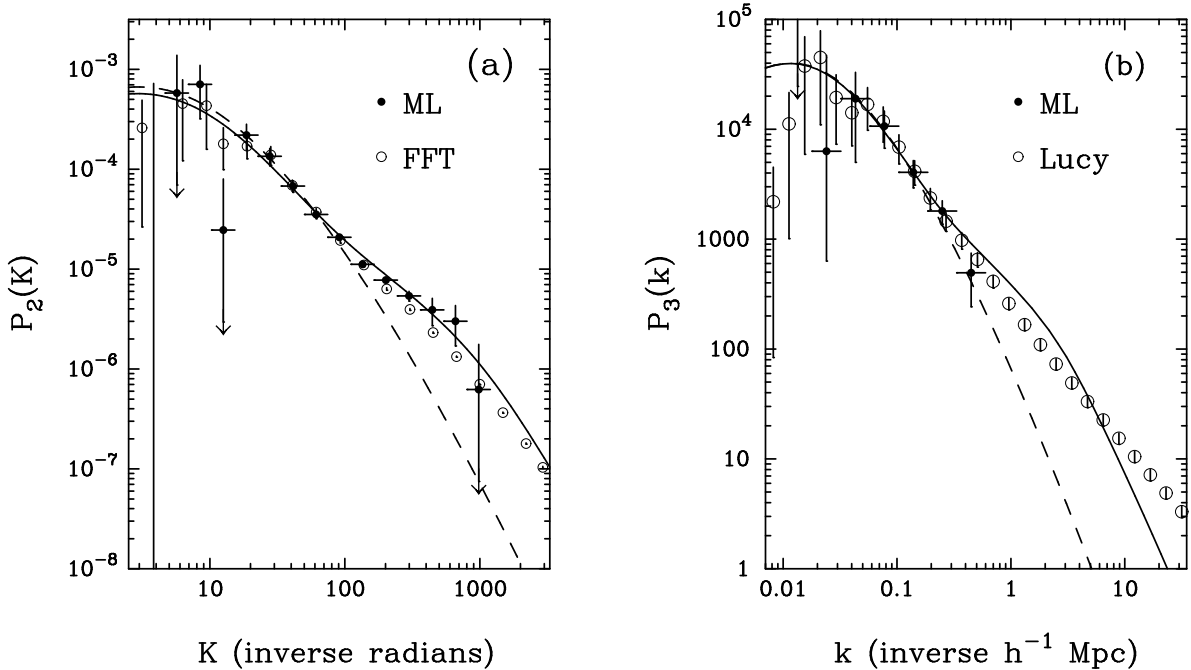


\vskip 3.6 truein

\includegraphics{pg2dfft.ps}
\includegraphics{pg3dfft.ps}

\caption
{The filled circles show the ML bandpower estimates of $P_2(K)$ and
$P_3(k)$ as plotted in Figure 6. The open circles in Figure (8a) show
the FFT estimates of $\bar P_2(K)$ and $1\sigma$ error bars.  The open
circles in Figure (8b) show the 3d power spectrum recovered by
applying the Lucy deconvolution algorithm described by BE94 to the FFT
$\bar P_2(K)$ estimates of Figure (8a). The error bars on the Lucy
$P_3(k)$ estimates show the inversions obtained by fitting to the tops
and bottoms of the error bars of the FFT $\bar P_2(K)$ estimates. The
solid line in Figure (8b) shows the scale-invariant CDM model,
including a non-linear correction as described in Section 4, that
provides a best fit to the maximum-likelihood points with $k < 0.33 h
{\rm Mpc}^{-1}$. The dashed line shows this model, but without the
non-linear correction. The solid and dashed lines in Figure (8a) show
the 2d power spectra computed from these fits to $P_3(k)$.}
\label{figure8}
\end{figure*}

In Figure (8a) we compare the ML 2d power spectrum
with the 2d power spectrum computed by applying an FFT to a $2048
\times 2048$ pixel  $b_J=20$ APM map. The FFT power spectrum is
computed as follows.
Let $\hat n({\bf K})$ be the Fourier transform of the observed counts
in cells of solid angle $\theta_c^2$,
\beglet
\begin{equation}
\hat n({\bf K}) = \sum_i n_i {\rm e}^{-i {\bf K}\cdot {\bf {x}_i}}, \label{A4a}
\end{equation}
and $\hat W({\bf K})$ be the Fourier transform of the survey window
function
\begin{equation}
\hat W({\bf K}) = \sum_i w_i {\rm e}^{-i {\bf K}\cdot {\bf {x}_i}}, \label{A4b}
\end{equation}
\endlet
where $w_i=1$ for active pixels and $w_i=0$ for masked pixels. We can form
an estimate of the 2d power spectrum from these Fourier transforms by
averaging
\begin{equation}
P_2({\bf K}) = {\vert \hat n({\bf K}) - {\cal N} \theta_c^2 \hat
W({\bf K}) \vert^2 - {\cal N} \Omega_s \over {\cal N}^2 \theta_c^2
\sum_{\bf K^\prime} \vert \hat W({\bf K} - {\bf K}^\prime)
\vert^2} \label{A5}
\end{equation}
over a range of wavenumbers centred on wavenumber $K$. (We will denote
this averaged estimate $\bar P_2(K)$). If the averaging is done over large
enough bins,  so that  estimates of $\bar P_2(K)$ in neighbouring
bins are weakly correlated, and the underlying fluctuations are assumed
to be Gaussian, then the variance of $\bar P_2(K)$ is given approximately
by 
\begin{equation}
{\rm Var}\; \left ( \bar P_2(K) \right)  = \left ( {N_T \over N_c} \right )
{1 \over N_K} \left [ {1 \over {\cal N}} + \bar P_2(K) \right]^2, \label{A6}
\end{equation}
where $N_T$ is the total number of pixels in the map, $N_c$ is the number
of active pixels and $N_K$ is the number of distinct wavenumbers used
to form the average $\bar P_2(K)$. (It is straightforward to derive
equation (\ref{A6}), using the approach of Feldman, Kaiser and Peacock
1994.)

\begin{figure*}
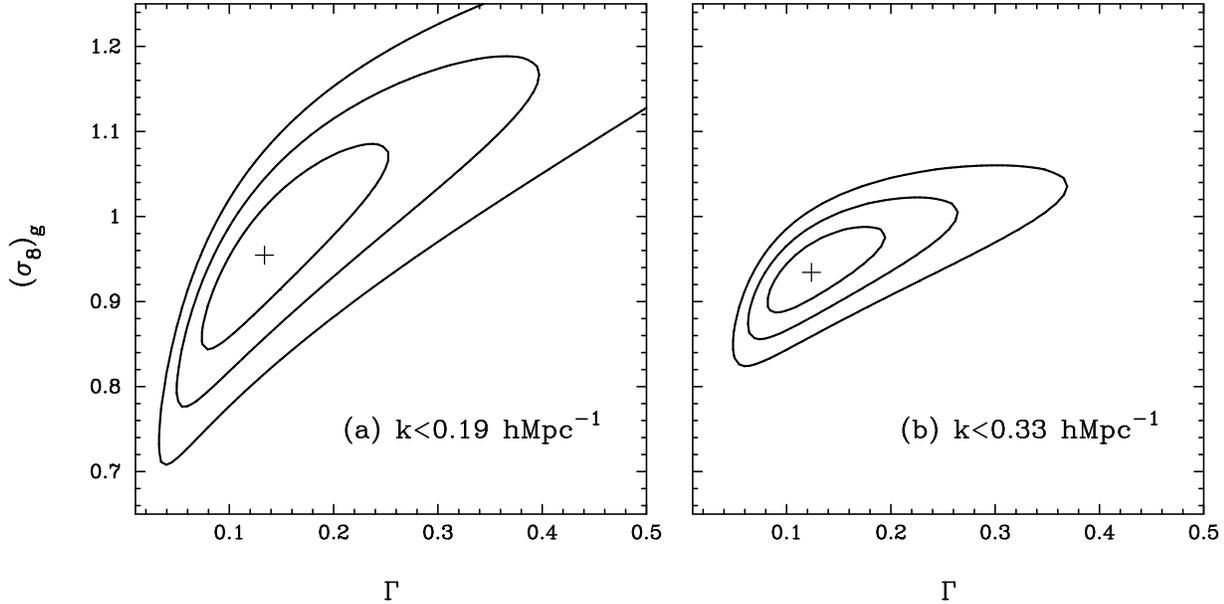


\vskip 3.3 truein

\includegraphics{pglike2pa.ps}
\includegraphics{pglike2pb.ps}

\caption
{One, two and three $\sigma$  contours for the amplitude
and CDM shape parameter $\Gamma$ determined from a likelihood fit
to the APM three dimensional power spectrum of Figure 6 for wavenumbers
(a)  $k \le 0.19 h{\rm Mpc}^{-1}$ and (b) $k \le 0.33 h{\rm Mpc}^{-1}$.
The crosses show the values of the parameters that minimise $\chi^2$.}
\label{figure9}
\end{figure*}

The FFT power spectrum points and error bars plotted in Figure (8a)
are computed from equations (\ref{A5}) and (\ref{A6}). These estimates
are not optimal for Gaussian random fields, unlike the ML estimates of
$P_2(K)$, and do not correctly deconvolve the survey window function
$\hat W({\bf K})$. Nevertheless, the FFT power spectrum agrees well
with the ML estimate, and even the error estimates
computed from equation (\ref{A6}) are in reasonable agreement with
those computed from the Fisher matrix. At wavenumbers $K \simgt 100$,
the error bars on the FFT estimates become smaller than the points on
the figure, whereas the error bars on the ML estimates
blow up. This is simply a consequence of the large pixel size
($\theta_c = 0.89^\circ$) of the map used for the ML
estimates compared to the pixel size ($\theta_c = 3.4^\prime$) of the
map used for the FFT estimates. As explained in Section 2.3, the
ML method is not guaranteed to be optimal or unbiased
at wavenumbers higher than $K \sim 100$, since the galaxy
distribution begins to show marked deviations from Gaussianity on
these scales. The FFT estimator will provide an unbiased estimate
at high wavenumbers, but the errors estimated from equation (\ref{A6})
will generally underestimate the true errors.

In Figure (8b) we show the ML inversion of $P_3(k)$
(filled circles) plotted against the Lucy inversion (open circles) of
the FFT $\bar P_2(K)$ estimates plotted in Figure (8a). The Lucy
inversion algorithm used here is exactly as described in BE94. The
error bars plotted on the open circles do not represent $1 \sigma$
error estimates, but simply indicate the ranges of the inversions
found by fitting to the tops and bottoms of the error bars of the FFT
points. The Lucy inversion is consistent with the ML
estimates and, of course, extend to much higher wavenumbers. What is
clear, however, is that the errors on the ML method at wavenumbers
$k \simlt 0.1 h {\rm Mpc}^{-1}$ are large and that there is
{\it no evidence for any turnover in the power spectrum} at smaller
wavenumbers. This agrees with the conclusions of Eisenstein and Zaldarriaga 
(1999). BE93, BE94, Maddox \etal~ (1996) claimed tentative evidence
for a turnover in the power spectrum based on an analysis of four
separate zones of the APM Survey. However, the scatter in the inverted
power spectra from the four zones is smaller than the Fisher matrix
errors computed from the ML inversion. The simulations on Gaussian
random fields described in Section (2.2) show that the Fisher matrix
error estimates are almost certainly the more reliable.

\section{Constraints on CDM models}

In this Section, we investigate the constraints on CDM models 
from the ML estimates of $P_3(k)$. Let $P^T(k)$ be a theoretical
model for the three-dimensional power spectrum specified by a
number of parameters. We find the parameters that minimise 
\begin{equation}
\chi^2 = \sum_{ij} F_{ij} \left (P_i - P^T(k_i) \right )\left
(P_j - P^T(k_j) \right ), \label{C1}
\end{equation}
where $F_{ij}$ is the Fisher matrix and $P_i$ the bandpower estimates
determined from the ML method.

We first investigate simple scale-invariant (scalar spectral index
$n_s = 1$) adiabatic CDM models. These models are characterized by 
a shape parameter $\Gamma$ and 
amplitude $(\sigma_8)_g$.
This
amplitude may differ from the amplitude of the mass fluctuations,
$(\sigma_8)_\rho$, depending on the relative bias between fluctuations
in the galaxy and the mass distributions. We also include a non-linear
correction to the shape of the power spectrum using the
formulae in Peacock and Dodds (1996). The non-linear correction
requires assumptions about the background cosmology and the amplitude
of the mass fluctuations. We fix the amplitude of the mass fluctuations
to that inferred by Eke, Cole and Frenk (1996) from the temperature
distribution of X-ray clusters:
$$
(\sigma_8)_\rho = 0.52 \Omega_m^{-0.52 + 0.13 \Omega_m}, \quad 
\Omega_m + \Omega_\Lambda = 1,
$$
and adopt  $\Omega_m = 0.3$, $\Omega_\Lambda=0.7$, consistent with the
latest results from CMB anisotropy measurements combined with
observations of distant Type Ia supernovae (see {\it e.g.}  de
Bernadis \etal, 2000; Hannay \etal, 2000; Efstathiou \etal, 1999).
The parameters $(\sigma_8)_\rho$, $\Omega_m$ and $\Omega_\Lambda$ are
kept fixed and we vary just the two parameters $(\sigma_8)_g$ and $\Gamma$ to
minimise $\chi^2$. The non-linear corrections become significant
only for $k \simgt 0.2  h {\rm Mpc}^{-1}$, thus provided 
that the sums
in equation (\ref{C1}) are restricted to low wavenumbers, the fits
will be insensitive to the non-linear model.

Likelihood contours for the two parameter fits are shown in Figures
(9a) and (9b). In Figure (9a), the fit is restricted to wavenumbers $k
< 0.19 h {\rm Mpc}^{-1}$ and in Figure (9b) it is resticted to $k <
0.33 h {\rm Mpc}^{-1}$ ({\it i.e.}  the first five and six points
plotted in Figure 6a respectively).  The values of $(\sigma_8)_g$ and $\Gamma$
that minimise $\chi^2$ are very similar in these two cases ($(\sigma_8)_g \approx
0.93$, $\Gamma \approx 0.12$), but the error contours are much smaller
for $k < 0.33 h {\rm Mpc}^{-1}$.  The best fitting model for $k < 0.33
h {\rm Mpc}^{-1}$ is plotted in Figure 8, which also illustrates the
size of the non-linear correction.  The error contours of Figure (9a)
are probably reasonable conservative limits on $(\sigma_8)_g$ and $\Gamma$ for
the APM Galaxy Survey. Although the constraints on the parameters are
tighter in Figure (9b), the wavenumber range is beginning to extend
into the range where the non-linear correction is becoming
important. 

It is not primarily the accuracy of the non-linear correction, or its
dependence on cosmological parameters, that makes us lean toward the
more conservative limits of Figure (9a). Rather, it is the assumption
that the galaxy power spectrum has exactly the same shape as that of
the mass distribution which  we feel is poorly justified, especially on
scales where the mass distribution is becoming non-linear. For
example, Benson \etal (2000), using plausible (but physically
uncertain) assumptions about galaxy formation applied to N-body
simulations of the dark matter distribution in a $\Lambda$-dominated
CDM model, find that the galaxy distribution displays {\it non-linear}
biasing on scales $\simlt 3 h^{-1} {\rm Mpc}$. In fact non-linear
biasing has been found in many investigations, including those based
on some of the earliest numerical simulations of the CDM model (Davis
\etal, 1985).  The nature of the bias depends on the physics of galaxy
formation and so is difficult to predict theoretically (see {\it e.g.}
Seljak 2000). We are therefore skeptical about using measurements of
the galaxy power spectrum together with, say, CMB anisotropy
measurements for the precise determination of cosmological parameters
(see {\it e.g.} Eisenstein, Hu and Tegmark 1998, 1999;
Wang, Spergel and Strauss 1999). As Figure 9
demonstrates, the likelihood constraints from galaxy clustering are
extremely sensitive to the range of wavenumbers used in fitting the
theoretical model or (almost equivalently) to the range of wavenumbers
over which galaxies are assumed to trace the mass distribution in some
simple way ({\it e.g.} constant bias).

 The constraints in Figure 9 are in qualitative agreement with previous
analyses of the APM Galaxy Survey ({\it e.g.} BE93, 94; Maddox \etal,  1996)
which  concluded that the large scale clustering of the APM Survey was
well fitted by a scale-invariant CDM model with $\Gamma \sim 0.2$. The 
contours in Figure 9a are, however, considerably tighter than the analogous
plot in Eisenstein and Zaldarriaga (1999, their Figure 3). Figure (9a) is
almost certainly more reliable because it is based on a self-contained
ML analysis of the APM map, rather than using estimates of $w(\theta)$
and its errors as an intermediate step, as in the
analysis of Eisenstein and Zaldarriaga. These authors argue
that their limits are close to the (approximate) theoretical lower
bounds on the parameters $(\sigma_8)_g$ and $\Gamma$ computed from the formula
\begin{equation}
\chi^2 = {\Omega_s \over 4 \pi}
\int K \left( {\delta P_2(K) \over P_2(K)} \right)^2 dK, \label{C2}
\end{equation}
where $\delta P_2(K)$ is the difference between the true 2d power
spectrum and a model with a different value of $(\sigma_8)_g$ and $\Gamma$.
While we agree with this equation, our evaluation of the constraints
on $(\sigma_8)_g$ and $\Gamma$ computed from this formula differ somewhat
from those of Eisenstein and Zaldarriaga.

\begin{figure}

\vskip 3.0 truein

\includegraphics{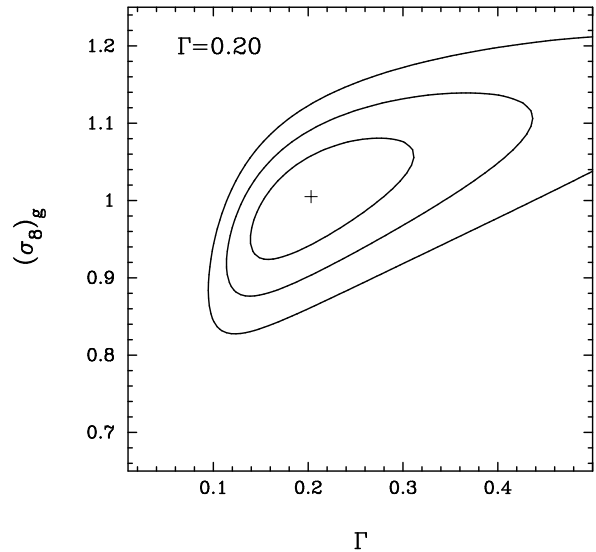}

\caption
{Lower bounds on on the amplitude and shape parameter $\Gamma$
computed from equation (26). We show $1$, $2$ and $3\sigma$ contours
for a  target model CDM model with $(\sigma_8)_g=1$ and $\Gamma = 0.20$.}
\label{figure10}
\end{figure}

\begin{figure*}
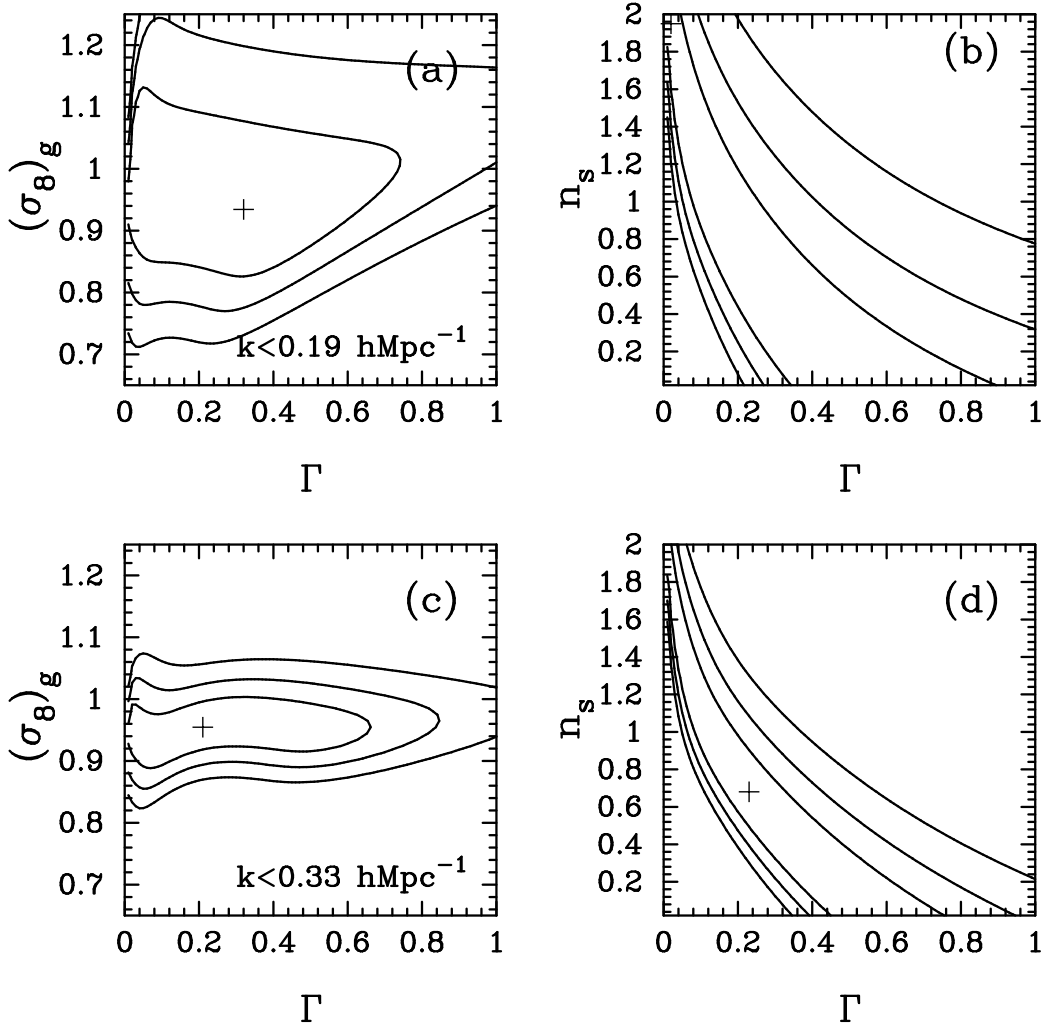


\vskip 5.8 truein

\includegraphics{pglike3pa.ps}

\includegraphics{pglike3pb.ps}

\caption
{Results of three parameter fits of CDM models to the APM 3d power
spectrum. Figures 11(a) and 11(c) show one, two and three $\sigma$
likelihood contours in the $(\sigma_8)_g$--$\Gamma$ plane after
marginalizing over $n_s$. Figures 11(b) and 11(d) show one, two and
three $\sigma$ likelihood contours in the $n_s$--$\Gamma$ plane after
marginalizing over $(\sigma_8)_g$. Figures 11(a) and (b) show results
for wavenumbers $k \le 0.19 h{\rm Mpc}^{-1}$ and Figures 11(c) and
11(d) show results for $k \le 0.33 h{\rm Mpc}^{-1}$.  The crosses
indicate the values of the parameters that maximise the marginalized
likelihoods.}
\label{figure11}
\end{figure*}

Our constraints are illustrated in Figure 10 for a target model
with $(\sigma_8)_g=1$ and $\Gamma=0.20$. As in Eisenstein and Zaldarriaga,
in computing $\delta P_2(K)$ from $P_3(k)$ we have set $P_3(k)$ equal
to zero for $k > 0.2  h {\rm Mpc}^{-1}$ so that wavenumbers
above this limit make no contribution to the $\chi^2$ 
in (\ref{C2}). Our analysis is consistent, in the sense that the
error contours in Figure 10 are tighter than those for the
real survey plotted in Figure (9a). It is not clear, however, 
why our evaluation of equation (\ref{C2})
differs from that of Eisenstein and Zaldarriaga.

We have also investigated three parameter fits to CDM models, varying
the scalar spectral index $n_s$ in addition to $(\sigma_8)_g$ and $\Gamma$. Figure 11 shows
likelihood contours in the $(\sigma_8)_g$--$\Gamma$ and $n_s$--$\Gamma$ planes
after marginalizing over the third parameter in each case assuming a
uniform prior. As in Figure 9, we have shown results for $k < 0.19 h
{\rm Mpc}^{-1}$ and $k < 0.33 h {\rm Mpc}^{-1}$ to demonstrate the
sensitivity of the results to the wavenumber ranges used in the fits.

Introducing $n_s$ as an additional parameter significantly weakens
the constraints on $(\sigma_8)_g$ and $\Gamma$. It is interesting that from
Figure (11b) we can infer, reasonably conservatively, that 
 a CDM model with $\Gamma \approx 0.5$ must
have a significant tilt of $n_s \simlt 0.8$ to be compatible
with the APM power spectrum on large scales.

\section{Conclusions}

In this paper we have tested and applied a maximum likelihood method
to estimate the 2d and 3d power spectra of the galaxy distribution
from a two-dimensional catalogue with a known redshift
distribution. The methods are similar to those applied to estimate the
power spectrum $C(\ell)$ of the CMB anisotropies. However, applied to
galaxy clustering, the ML method provides an optimal way of estimating
the three-dimensional power spectrum and its covariance matrix
directly from the 2d data. This provides a simple alternative to
inverting the integral equations relating the 2d power spectrum, or
angular correlation function, to the 3d power spectrum.

We have investigated the effects of Galactic extinction on the APM
survey using the extinction maps of Schlegel, Finkbeiner and Davis
(1998). Galactic extinction is shown to have little  effect 
on the power spectrum over the APM area with $\delta < - 20^\circ$.
Eliminating the small regions below this declination limit where
the extinction exceeds $0.2$ magnitudes depresses the power spectrum
of dust still further so that it has a neglible effect on the power
spectrum of galaxy clustering.

Our results show that the ML power spectra are in good agreement with
previous estimates of the 2d and 3d power spectra of the APM Survey
(BE93, BE94; Maddox \etal, 1996; Eisenstein and Zaldarriaga, 1999;
Dodelson and Gazta\~naga 2000). The ML method produces reliable
estimates of the covariance matrices of the 2d power spectra.  In
agreement with Eisenstein and Zaldarriaga, we conclude that the errors
on the 3d power spectrum have been underestimated in earlier papers
(BE93, BE94) and that there is no evidence for a peak, or turnover, in
the APM galaxy power spectrum at $k \simlt 0.1 h {\rm Mpc}^{-1}$. By
fitting a scale invariant CDM model to the 3d power spectrum at
wavenumbers $k \le 0.19 h {\rm Mpc}^{-1}$ we find the amplitude and
shape parameter lie within the ranges $0.78 \le (\sigma_8)_g \le 1.18$
and $0.05 \le \Gamma \le 0.38$ at the $2\sigma$ level. Including the
scalar spectral index $n_s$ as a parameter significantly weakens the
constrains. Nevertheless, compatibility with the APM data requires
that CDM models with $\Gamma \simgt 0.5$ have spectral indices 
$n_s \simlt 0.8$ at the $2\sigma$ level.

The methods described here have applications to other 2d surveys, for
example, the forthcoming Sloan Digital Sky Survey (SDSS, see {\it
e.g.}  Margon 1999). It is also possible that a pixel based ML method,
as described here, may prove useful for the analysis of 3d galaxy
redshift surveys such as the 2dFGRS and the SDSS. The likelihood 
distributions derived here may have some applications in parameter
estimation studies using CMB and other data ({\it e.g.} Jaffe \etal, 2000;
Lange \etal, 2000). The likelihood distributions plotted in Figures 9
and 11 are available from the authors on request.

\bigskip

\noindent
{\bf Aknowledgements:} S.J. Moody thanks PPARC for the award of a research
studentship. We also thank the 2dFGRS team for allowing us to use
some of their data prior to publication.

\end{document}